\documentclass[twocolumn,showpacs,preprintnumbers,amsmath,amssymb]{revtex4}

\usepackage{graphicx}
\usepackage{dcolumn}
\usepackage{bm}

\begin{document}

\title [] {Unconventional superconductivity in weakly correlated,
non-centrosymmetric $\rm{ Mo_3Al_2C}$}

\author{E. Bauer,$^1$ G. Rogl,$^2$  Xing-Qiu Chen,$^3$ 
R.T. Khan,$^1$ H. Michor,$^1$ 
G. Hilscher,$^1$ E. Royanian,$^1$ K. Kumagai,$^4$ D.Z. Li,$^3$ Y.Y. Li,$^3$  
 R. Podloucky,$^2$ and P. Rogl$^2$}
\affiliation{
$^1$Institute of Solid State Physics, Vienna University 
of Technology, A-1040 Wien, Austria 
\\
$^2$Institute of Physical Chemistry, University of Vienna,
A-1090 Wien, Austria\\
$^3$Shenyang National Laboratory for Materials Science, 
Institute of Metal Research, Chinese Academy of Sciences, Shenyang, China \\
$^4$Division of Physics, Graduate School of Science,
Hokkaido University, Sapporo, 060-0810, Japan
}

\date{\today}

\begin{abstract}

Electrical resistivity, specific heat and NMR measurements 
classify non-centrosymmetric $\rm Mo_3Al_2C$ 
($\beta$-Mn type, space group $P4_132$)
as a strong-coupled superconductor with $T_c = 9$~K deviating
notably from BCS-like behaviour. 
The absence of a Hebbel-Slichter peak, a power law behaviour
of the spin-lattice  relaxation rate
(from $^{27}$Al NMR), a $T^3$ temperature dependence of the specific 
heat and a pressure enhanced $T_c$ suggest unconventional superconductivity
with a nodal structure of the superconducting gap.
Relativistic DFT calculations reveal a splitting of 
degenerate electronic bands  due to the asymmetric spin-orbit coupling,
favouring a mix of spin-singlet and spin triplet components in the 
superconducting condensate, in absence of strong correlations 
among electrons.

\end{abstract}


\maketitle


Carbides based on Mo comprise a large body of refractory 
compounds,  where carbon atoms 
(in trigonal prismatic or octahedral $\rm Mo_6C$ subunits) 
occupy a fraction of the interstitial
sites either in an ordered or in a random manner. 
Among Mo-based carbides for which superconductivity (SC)
was reported ($\alpha$MoC at $T_c = 9.95$~K,
$\eta$MoC at 7.57~K, $\rm Mo_2BC$ at 6.33~K and $\rm Mo_3Al_2C$ at 9.05~K)
the crystal structure of $\rm Mo_3Al_2C$
is outstanding, since the respective $\beta$-Mn type does not possess
a center of inversion \cite{Toth1968}. The missing
inversion symmetry might initiate 
a mixture of spin-singlet and spin-triplet pairs in the
SC condensate \cite{Gorkov} as was recently proposed to
explain SC in $\rm CePt_3Si$ \cite{Bauer2004},
$\rm UIr$ \cite{Akazawa}, $\rm CeRhSi_3$
\cite{Kimura}, and $\rm CeIrSi_3$ \cite{Sugitani}.
Non-centrosymmetry (NCS) of the crystal structure introduces an electrical field
gradient and, thereby,
creates a Rashba-type antisymmetric spin-orbit coupling \cite{Gorkov}.

The Ce and U-based SCs  indicated above are characterised by heavy
fermion behaviour at low temperatures provoked by 
Kondo interaction.
NCS in such systems can lead to new anomalous spin fluctuations, 
stabilizing triplet pairing, in addition to the singlet part
\cite{Takimoto2009}.
On the other hand, a variety of SCs has been identified, 
which lacks strong electron correlations as well as
a centre of inversion.
For a recent listing of these systems
see Ref. \cite{Bauer2009}. Except $\rm Li_2Pt_3B$ \cite{Yuan},
all yet studied NCS SCs without
strong correlations among electrons are typical
$s$-wave fully gapped BCS SC either weakly or strongly coupled.

In order to shed light onto the primary mechanism
activating unconventional SC, we are searching
for systems where SC occurs in absence of
inversion symmetry, and also in absence of strong electron
correlations. Revisiting
$\rm Mo_3Al_2C$ ($\beta$-Mn structure), we aim to
extend research done in the 1960's \cite{Toth_book}, providing
insight into microscopic features and the 
electronic structure.


For the preparation of $\rm Mo_3Al_2C$ an elemental powder mixture
(purity $>$ 99.9 mass\%, about
5 g) was cold compacted, reacted in a high 
vacuum furnace for 24 hrs at $1500^{\circ}$C
with one intermediate grinding and compacting step. 
Afterwards the material was ball milled
and hot pressed at $1250^{\circ}$C at 56 MPa.
 Refinement of the crystal structures was performed
with the program Fullprof \cite{Roisnel}.  Measurements of
physical properties were carried out with 
standard techniques \cite{Bauer3,Bauer100}. 
The density functional theory (DFT) 
calculations were performed with the {\em
Vienna ab initio Simulation Package} (VASP) \cite{vasp_theory}. 
For details see our recent paper on NCS $\rm BaPtSi_3$ \cite{Bauer2009}.

\begin{figure}[b]
\begin{center}
		\includegraphics[width=0.4\textwidth]{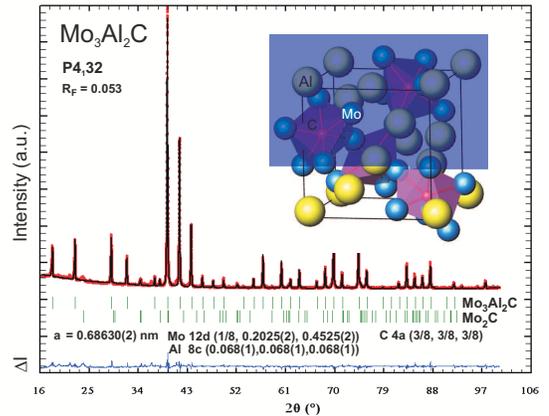}
\end{center}
\caption{Rietveld refinement (Guinier-Huber image plate system, $CuK_{\alpha1}$) 
and crystallographic data of $\rm Mo_3Al_2C$. The inset 
shows a 3-dimensional view 
of the crystal structure. Traces of $\rm Mo_2C$ are indicated by vertical bars.
}
\label{fig1}
\end{figure}

X-ray Rietveld refinement confirmed a cubic, 
non-centrosymmetric structure (space group $P4_132$),
isotypic to the $\beta$-Mn type; see Fig. \ref{fig1}).

\begin{figure}[]
\begin{center}
		\includegraphics[width=0.4\textwidth]{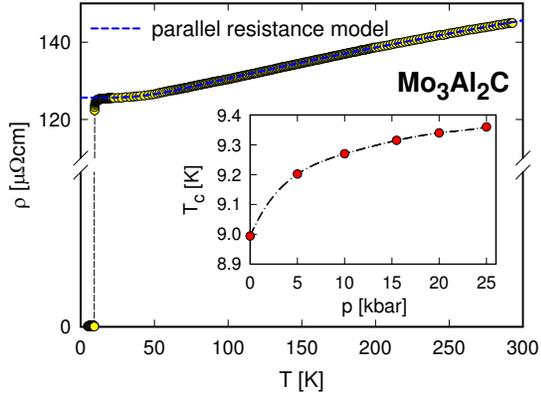}
\end{center}
\caption{(Color online)
Temperature dependent electrical resistivity
$\rho$ of $\rm Mo_3Al_2C$.
The dashed line is a least squares fit according to
the parallel resistance model. 
The inset shows the pressure dependence of $T_c$.
}
\label{fig_resi}
\end{figure}

Measurements of the temperature dependent 
electrical resistivity $\rho$ of $\rm Mo_3Al_2C$
clearly evidences metallic behaviour and 
indicate a SC phase transition at $T_c = 9$~K (see Fig. \ref{fig_resi}),
in agreement with the data reported by
Johnston et al. \cite{Johnston64}. SC with almost 100~\%
volume fraction is revealed from magnetic susceptibility
measurements as well.
Since the absolute resistivity values are large,
the parallel resistance model (compare e.g., Ref. \cite{Gunnarsson}) can be used
to describe $\rho(T)$, where the ideal resistivity
follows from the Bloch-Gr\"uneisen model.
A fit employing this model is shown in Fig. \ref{fig_resi}
as a solid line, revealing a Debye temperature $\theta_D = 286$~K and a saturation
value $\rho_{sat} = 350~\mu \Omega$cm. 
An estimation of the electron-phonon
interaction  strength $\lambda_{e,ph}$ is possible in terms of the
McMillan formula \cite{McMillan}.
Applying this model, and 
taking the repulsive screened Coulomb part
$\mu^*\approx 0.13$,
yields  $\lambda_{e,ph} \approx 0.8$;
this characterizes $\rm Mo_3Al_2C$ as 
a SC well beyond the weak coupling limit.

The pressure dependence of $T_c$ of $\rm Mo_3Al_2C$ is displayed in the 
inset of Fig. \ref{fig_resi}. Obviously, $T_c(p)$ increases,
but tends to saturate for high pressures. An increase of $T_c$ 
is rarely found in a simple materials; rather, such a behaviour 
frequently occurs in unconventional SCs like in 
high temperature SCs, in various pyrochlores, in some Fe-pnictides or heavy
fermion materials. 
Bogolyobov et al. \cite{Bogo} demonstrated that there are 
two principal parameters determining $T_c$: $\theta_D$ and 
the electronic density of states at the Fermi energy, $N(E_F)$. 
Since the application of pressure hardly modifies
$\rho(T,p)$ in the normal state region
(not shown here),  $\theta_D(p)$ remains
unchanged. Thus, a slight increase of $N(E_F)$ is 
concluded, enhancing $T_c$
on pressurizing  $\rm Mo_3Al_2C$.

\begin{figure}[t]
\centering
		\includegraphics[width=0.48\textwidth]{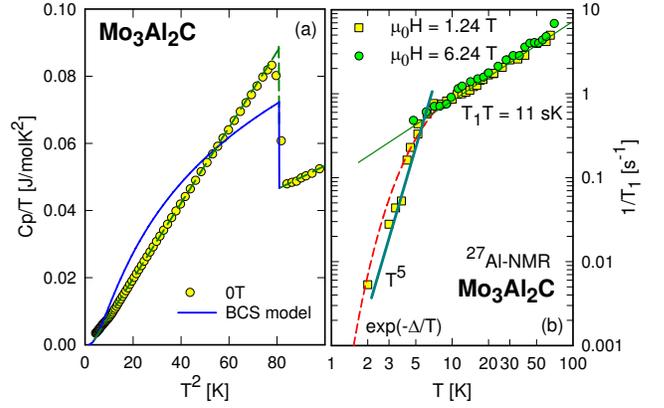}
\caption{ (Color online) (a) Temperature dependent specific heat $C_p$ of
$\rm Mo_3Al_2C$ plotted as $C_p/T$ vs. $T^2$. The dashed line
is a guide for the eye and
indicates an idealised superconducting phase transition
together with a $T^3$ dependence of $C_p(T)$ 
for $T < T_c$.  The solid line represents $C_p(T)$
of a spin-singlet fully gapped BCS 
superconductor according to M\"uhlschlegel \cite{Muehl}.
(b) Temperature dependent $1/T_1$ $^{27}$Al relaxation rate 
deduced at $\mu_0H = 1.24$ and 6.95~T. 
The solid lines are guides for the eye.
The dashed line represents an exponential temperature dependence.}
\label{fig3}
\end{figure}

Fig. \ref{fig3} shows the temperature dependent specific  heat $C_p$ of
$\rm Mo_3Al_2C$ taken at 0 T and plotted as $C_P/T$ vs. $T^2$.
Bulk SC is evidenced from
a distinct anomaly at 9~K, rendering the onset of the SC
phase transition. 
A closer inspection of the data gives evidence of
various non-BCS like features: i) The jump of the specific heat
at $T_c$, $\Delta C_p/(\gamma_n T_c) \approx 2.28$,
is well above the value expected for an $s$-wave
BCS SC with $\Delta C_p/(\gamma T_c) \approx 1.43$.
This clearly evidences strong-coupling SC. 
ii) The temperature
dependent heat capacity below $T_c$ significantly deviates from
the universal BCS dependence as 
indicated by the solid line. Rather, a power law with $C_p (T < T_c) \propto T^3$
is obvious from the experimental data 
(compare Fig. \ref{fig3}), which is sketched by
the dashed line as well. Such a temperature dependence excludes a fully gapped
SC state; instead, a nodal structure is likely, 
where the SC gap vanishes along points.

\begin{table}[b]
        \caption{Normal state and SC properties of $\rm Mo_3Al_2C$.}
        \begin{tabular}{|c|c|}   \hline \hline
        crystal structure             &   cubic, $\beta$-Mn type \\
        space group                   &   $P4_132$        \\
        lattice parameter             &   $a = 0.68630$~nm \\
\hline
        Sommerfeld value              &   $\gamma_n = 17.8$~mJ/molK$^2$ \\
        Debye temperature             &   $\theta_D = 315$~K  \\ 
        transport mean free path      &   $ \lambda _{tr} = 3.06$~nm  \\ 
\hline
        transition temperature         &   $T_{c} = 9.0$~K \\
        electron-phonon enhancement factor  &   $1 + \lambda_{e,ph} = 1.8$ \\
        upper critical field           &   $\mu_0H_{c2}(0) \approx 15.7$~T \\
        slope of upper critical field  &   $\mu_0H_{c2}' = -3$~T/K    \\
        thermodynamic critical field   &   $\mu_0H_c(0) =  0.146$~T \\
        correlation length             &   $\xi \simeq 4.6$~nm    \\
        Ginzburg Landau parameter      &   $\kappa_{GL} \simeq 76$  \\
        London penetration depth       &   $\lambda \simeq 380$~nm  \\
        nodal structure                &   point-nodes   \\
\hline \hline
        \end{tabular}
    \label{table1}
\end{table}

The $1/T_1$ $^{27}$Al relaxation rate, taken at $\mu_0H = 1.24$~T 
and partially at 6.95~T 
is plotted in Fig. \ref{fig3}(b) on a double logarithmic scale.  
A Hebbel-Slichter peak right at $T_c$ is absent. This 
is compatible with a partial disappearance of 
the SC gap at the Fermi energy,
in line with non-$s$-wave SC. Below $T_c$, 
a non-exponential but rather a $T^n$ 
temperature dependence hints towards a nodal structure, 
closing partially the SC gap at the Fermi surface. 
We note that a 
$1/T_1 \propto T$ component, expected as a signature
of a finite impurity density of states, is clearly absent in our  
low temperature data.
Volovik and Gork'ov  \cite{Volo1985}
have shown that a proportionality of the 
density of states according to $N(E) \propto E^{m}$ results in a NMR 
relaxation rate $1/T_1 \propto T^{2m+1}$. Thus, an anisotropic gap with nodal
structures yields, in general, a $T^n$ 
power law of $1/T_1$ with $n = 3$ for line nodes
and $n = 5$ for point nodes. Intersecting nodes, however, might modify such simple
temperature dependencies \cite{Hasegawa1996}. 
Furthermore, Hayashi et al. \cite{Hyashi} 
demonstrated that NCS SCs with  mixed spin singlet and triplet states infer
a rather unconventional $1/T_1$ relaxation rate.

\begin{figure}[t]
\centering
		\includegraphics[width=0.48\textwidth]{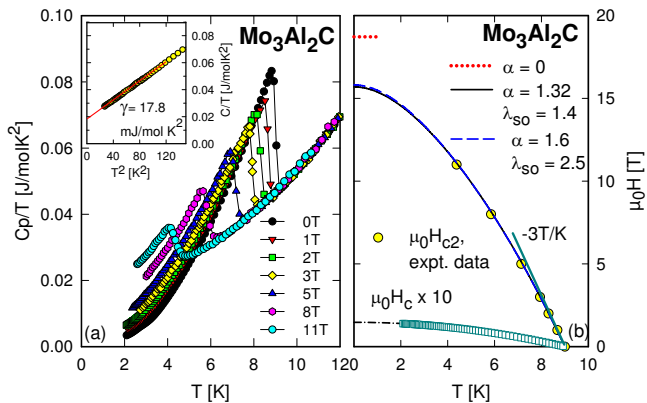}
\caption{(Color online) (a) Temperature and field dependent specific heat
$C_p$ of $\rm Mo_3Al_2C$. (b)  Temperature dependent upper critical field
	 $\mu_0 H_{c2}$ and thermodynamic critical field
	 $\mu_0 H_{c}$ as obtained from
	 specific heat measurements. The solid and the  long-dashed
	 lines are fits according to the WHH model for different
	 values of the Maki parameter. The horizontal bar indicates the upper
	 critical field $\mu_0H^*(0)$ in absence of Pauli-limiting. 
	 The dashed-dotted line is an extrapolation
	 of the thermodynamic critical field towards zero.
}
\label{fig33}
\end{figure}

Summarized in Fig. \ref{fig33}(a) is the field and temperature dependent heat
capacity of $\rm Mo_3Al_2C$, highlighting the suppression of SC
upon the application of a magnetic field. 
The fact that even fields of 11 T do not
suppress superconductivity evidences a large upper critical field
$\mu_0 H_{c2}$ as well as a large initial slope $\mu_0 H_{c2}'$.
The extension of the normal state behaviour towards lower temperatures with rising
magnetic fields allows to obtain in a standard manner the Sommerfeld value 
$\gamma = 17.8$~mJ/molK$^2$
and $\theta_D \approx 315$~K (compare Fig. \ref{fig33}(a), inset).
The accurate determination of $\gamma$ and of $T_c(\mu_0H)$ 
was accomplished by idealizing the heat capacity anomaly under the constraint
of entropy balance between the superconducting and the normal state.
$T_c(\mu_0H)$ obtained from Fig. \ref{fig33}(a)
is plotted in Fig. \ref{fig33}(b).

The temperature dependency of
$\mu_0 H_{c2}$ is described following 
the model of Werthammer et al. \cite{Werthamer},
incorporating orbital pair-breaking, the effect
of Pauli spin paramagnetism and spin-orbit scattering.
Two parameters, the Maki parameter 
$\alpha$ (Pauli paramagnetic limitation) \cite{Maki}
and spin-orbit scattering $\lambda_{so}$ specify this model.
While an increase of $\alpha$ decrements the upper critical field,
an increase of $\lambda_{so}$ compensates the former, restoring
for $\lambda_{so} \rightarrow \infty$ a maximum field constrained
from orbital pair breaking only.
In a first
approximation, the Maki parameter $\alpha$ can be derived   from
$\gamma$ and $\rho_0$ 
\cite{Werthamer}, resulting in $\alpha = 1.32$. Alternatively,
$\alpha$ can be estimated from $\mu_0 H_{c2}'$ \cite{Maki}, 
 revealing $\alpha^* = 1.6$.
The sizable Maki parameter of both approximations
is an indication that
Pauli limiting is non-negligible in $\rm Mo_3Al_2C$.

\begin{figure}[b]
\centering
		\includegraphics[width=0.4\textwidth]{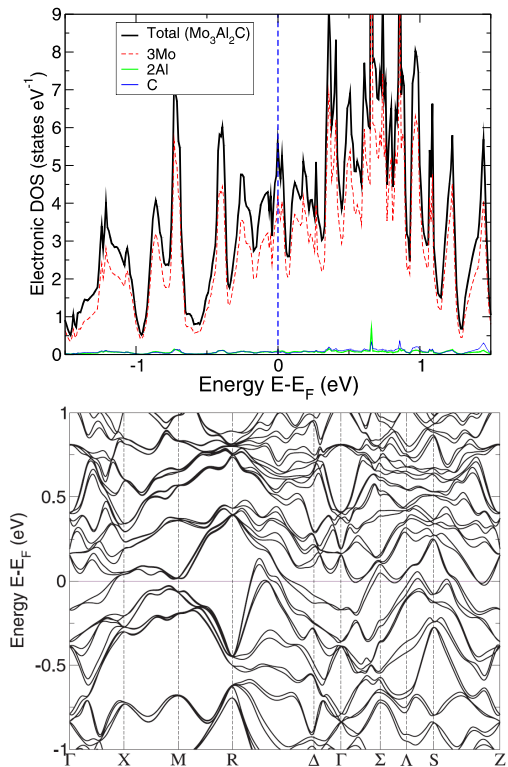}
\caption{(Color online)
(Upper panel)
Section of relativistic total and atom-projected densities of states
(DOS) in states eV$^{-1}$ for $\rm Mo_3Al_2C$ 
summed over all three Mo atoms for 
the energy range $\pm 1.5$ around the Fermi energy $E_F$.
(Lower panel): Relativistic electronic band structure 
along high symmetry directions for
$\rm Mo_3Al_2C$ in the energy range $\pm 1.$~eV around the Fermi energy $E_F$.
}
\label{fig:dos_theory}
\end{figure}

Using $\alpha = 1.32$  ($\alpha^* = 1.6$)
and $\mu_0 H_{c2}' = - 3$~T/K
yields $\mu_0 H_{c2}(T)$  as displayed
as solid and dashed lines in Fig. \ref{fig33}(b) for 
$\lambda_{so} = 1.4$ and $\lambda_{so} = 2.5$, respectively,
with $\mu_0 H_{c2} (0) \approx 15.7$~T.
The Pauli limiting field follows from 
$\mu_0 H_p(0) = \sqrt 2 \mu_0 H^*(0) / \alpha $, 
where $\mu_0 H^*(0) = 18.72$~T, is the 
WHH result for $\alpha = 0$, i.e., the orbital limit.
Thus, $\mu_0 H_p(0) = 20$~T for the former and 16.5~T for the latter value of
$\alpha$. These values are in line with $\mu_0 H_p(1.2~{\rm K}) = 15.6$~T
reported by Fink et al. \cite{Fink}. 
In the case of strong coupling superconductivity, these values are further
enhanced according to $H_p^{str}(0) = H_p(0) (1+\lambda_{e,ph})^{\epsilon}$ with 
$\epsilon = 0.5$ or 1.0 \cite{Orlando,Schossmann}. Hence, 
Pauli limiting is not the principal mechanism restricting the upper critical field
in $\rm Mo_3Al_2C$, but is present in a relevant size.

The thermodynamic critical field $\mu_0 H_c(T)$ derived 
from heat capacity data (compare e.g. Ref. \cite{Bauer2009})
is shown in Fig. \ref{fig33}(b) by open squares;
an extrapolation to $T \rightarrow 0$ (dashed-dotted line)
yields $\mu_0H_c(0) \approx ~0.146$\,T.

SC and normal state parameters of $\rm Mo_3Al_2C$ 
can be assessed from $\gamma$,
$\mu_0 H_{c2}'$, $\mu_0 H_{c2}(0)$ and $\rho_0$ \cite{Orlando,Tinkham}.
From the Ginzburg Landau theory with the thermodynamic 
critical field as primary input,  
the coherence length, the Ginzburg Landau parameter and the London 
penetration depth are calculated. Parameters  are 
summarized in Table \ref{table1}. Based on the estimate
$l_{tr}/\xi \approx 0.66$ we classify  $\rm Mo_3Al_2C$ as
a superconductor in the dirty
limit; $\kappa_{GL} \approx 76$ refers to a type II superconductor.

A section of the
calculated electronic density of states (DOS) 
of $\rm Mo_3Al_2C$  is shown in
Fig. \ref{fig:dos_theory}  for a relativistic  
calculation including spin orbit coupling
(upper panel).  The DOS around the Fermi energy
stems  primarily from from Mo-$4d$ states, whilst the contribution of
Al and C is almost negligible. The low partial Al DOS calculated
at $E_F$ corresponds well to the 
NMR Korringa constant, $T_1T = 11$~sK
($1/T_1T \propto N(E_F)^2$). 
A comparison with Al metal ($T_1T = 1.8$~sK) reveals a local Al DOS in
$\rm Mo_3Al_2C$ of about  3~\%  with respect to the total DFT DOS.

The Fermi energy $E_F$ of $\rm Mo_3Al_2C$
is located in a local maximum of the DOS;
its large value favours SC. 
Employing the Sommerfeld expansion,
$N(E_F) = 5.48$ states/eV corresponds to $\gamma_b = 12.9$~mJ/molK$^2$,
in fair agreement with $\gamma = \gamma_b(1+\lambda_{e,ph}) = 17.8$~mJ/molK$^2$.

The lower panel of Fig. \ref{fig:dos_theory} 
displays the DFT electronic band structures
along high symmetry directions for $\rm Mo_3Al_2C$.
With respect
to a non-relativistic calculation (not shown here),
the degenerate bands become split due to the lack  of
inversion symmetry in  $\rm Mo_3Al_2C$.
Specifically, for all bands crossing the Fermi
energy the degeneracy is lifted, separating spin-up and spin-down
electrons. This provides conditions for the occurrence
of spin-singlet and spin-triplet 
Cooper pairs, leading to two gap functions, 
where each gap is defined on one of the two bands formed 
by degeneracy lifting. Superposition of these
gaps is presumed to constitute a nodal structure 
of the resulting SC gap as corroborated from 
the present experimental data.

In conclusion, 
electrical resistivity, specific heat and NMR measurements 
classify non-centrosymmetric $\rm Mo_3Al_2C$ 
as a strong-coupled SC with $T_c = 9$~K.
The temperature dependent specific heat and the $1/T_1$ 
$^{27}$Al NMR relaxation rate  
deviate from BCS predictions, thus referring to
a nodal structure of the superconducting
gap even though SC of $\rm Mo_3Al_2C$ occurs in the dirtly limit.
This manifests a robustness of the unconventional order parameter
of the NCS superconductor. Moreover,
unconventional pairing is in line with the splitting
of electronic bands  due to the asymmetric spin-orbit coupling
as revealed from relativistic DFT calculations.
These split bands might be the cause of a mixing of
spin-singlet and spin-triplet Cooper pairs, which
otherwise are distinguished by parity \cite{Gorkov}, 
making a nodal structure likely \cite{Frigeri,Hyashi}. Whilst this 
proposition has been corroborated for SCs with strong 
correlations among electrons, specific heat data  
unambiguously disprove 
a strongly correlated electronic state in $\rm Mo_3Al_2C$.  
In spite of a lack of correlations, unconventional SC seems
to arise from a substantial band splitting and the fact that
inversion symmetry is missing in {\em all} crystallographic directions. 
In these respects, $\rm Mo_3Al_2C$ is the only example
besides isomorphous $\rm Li_2Pt_3B$ \cite{Yuan}.

Work supported
by the Austrian Science Foundation FWF P22295.
X.-Q.C acknowledges the support from the ``Hundred Talents
Project'' of CAS.

\end{document}